\documentclass[a4paper]{article}
\usepackage{Odyssey2024}
\usepackage{epsfig,amssymb,amsmath}
\usepackage{booktabs}
\usepackage[verbose]{placeins}
\usepackage{caption}
\ninept

\renewcommand{\paragraph}[1]{\noindent\textbf{#1}\quad}
\usepackage{hyperref}
\usepackage{enumitem}

\title{Unraveling Adversarial Examples against Speaker Identification - Techniques for Attack Detection and Victim Model Classification}

\name{Sonal Joshi$^{*}$, Thomas Thebaud$^{*}$, Jes\'us Villalba, Najim Dehak}

\address{Center for Language and Speech Processing \\ Johns Hopkins University, Baltimore, MD, USA \\
{\small \tt sjoshi12@jhu.edu, tthebau1@jhu.edu} }
\begin{document}
\maketitle

\begin{abstract}
Adversarial examples have proven to threaten speaker identification systems, and several countermeasures against them have been proposed. In this paper, we propose a method to detect the presence of adversarial examples, i.e., a binary classifier distinguishing between benign and adversarial examples. We build upon and extend previous work on attack type classification by exploring new architectures. Additionally, we introduce a method for identifying the victim model on which the adversarial attack is carried out. To achieve this, we generate a new dataset containing multiple attacks performed against various victim models. We achieve an AUC of 0.982 for attack detection, with no more than a 0.03 drop in performance for unknown attacks. Our attack classification accuracy (excluding benign) reaches 86.48\% across eight attack types using our LightResNet34 architecture, while our victim model classification accuracy reaches 72.28\% across four victim models.

\end{abstract}

\section{Introduction}
Neural networks have revolutionized various domains, exhibiting remarkable performance for a range of speech related tasks, such as automatic speech recognition~\cite{alharbi2021automatic} or speaker recognition~\cite{snyder2018x}. 
However, the susceptibility of these networks to adversarial attacks~\cite{PGD, FGSM, CW} has raised significant concerns regarding their robustness and reliability~\cite{chen2022towards,wang2022survey}. 
Adversarial attacks exploit vulnerabilities in neural networks, causing misclassification or erroneous behavior due to (usually) small imperceptible perturbations to the input data. 
Applied to speech, this takes the form of an (almost) imperceptible noise that can drastically change the outputs of a given system. These attacks can be either \textit{white-box attacks} (when the network model parameters are available to the adversary) or \textit{black-box attacks} (when the network model parameters are not available to the adversary). Usually, white-box attacks are more powerful and take comparatively shorter time to generate than black-box attacks which rely on querying the model multiple times to generate an attack. With the widespread availability of pre-trained models, white-box attacks are feasible and increase the threat they represent.
In this work, we focus on speaker recognition systems. 
Concerned with the threat of these systems, a range of countermeasures~\cite{advreview,advSurvey,abdullah2021sok} have emerged. These countermeasures can be classified into two broad categories - Proactive defenses that make the models more robust and Reactive countermeasures that filter out adversarial attacks. 
\cite{villalba2020x,villalba2021representation} introduced a novel countermeasure strategy designed to classify between different attacks, applied on speaker identification and verification systems.

In this paper, we extend this work by improving multiple parts of the classification system.
Our proposed approach leverages insights from adversarial attack mechanisms to develop a comprehensive defense framework. Through extensive experimentation and analysis, we demonstrate the efficacy of our method in accurately detecting the presence of an attack and classifying the type of adversarial attack.
Additionally, we propose a method to identify the victim model on which the attack was performed. Toward this end, we generate a new dataset of attacks computed on a range of speaker identification networks.
By bridging the gap between attack detection and attack classification, our work represents a significant step towards building more resilient and trustworthy machine learning systems in the face of evolving adversarial threats.

Our main contributions can be summarized as:
\begin{itemize}[noitemsep,topsep=0pt]
   \item Development of a  binary adversarial attack detection system, complemented by an ablation study that explores the impact of omitting different types of adversarial attacks during training.
    \item Exploration of new architectures for attack classification system.
    \item Introduction of a new dataset featuring different adversarial attacks targeting four different victim models.
    \item Utilization of the above dataset to classify amongst victim models
\end{itemize}

Section \ref{sec:related_works} will present in detail the previous classification system this work is based on, as well as the adversarial attacks we choose to focus on.
Then, Section \ref{sec:datasets} presents the dataset used for all our experiments, followed by Section \ref{sec:threat_model}, presenting the various threat models that will be studied here: what attacks, norms, victim models will be considered, and the building of the datasets that contains all those threat models.
Section \ref{sec:experiments} presents our classification and detection experiments, which results are shown in Section \ref{sec:results}, then discussed in Section \ref{sec:conclusion}


\section{Related Work}
\label{sec:related_works}
This section presents previous works related to adversarial attack detection and classification.

\subsection{Adversarial attacks}
The range of existing adversarial attacks is increasingly expanding, however in this paper, we choose to focus on a set of existing un-targeted white-box adversarial attacks that has been tested against speaker recognition and verification systems~\cite{villalba2020x,villalba2021representation}. The reason for this choice is that white-box attacks are more powerful and take comparatively shorter time to generate than black-box attacks which rely on querying the model multiple times to generate an attack. Since, our focus is on attack classification, the compute would be significant.
We focus on three adversarial attacks: the Fast Gradient Sign Method (FGSM), Projected Gradient Descent (PGD) and Carlini-Wagner attack (CW).

\paragraph{Fast Gradient Sign Method}\cite{FGSM} was the first adversarial attack proposed, simply projecting an input toward the direction of the highest gradients to get the most impact on the outputs from the least modification of the inputs. Iterative-FGSM(iter-fgsm) extends the basic FGSM attack by applying the FGSM iteratively. Instead of crafting a single perturbation, the input is perturbed multiple times in small steps, and the gradient is computed again based on the perturbed input from the previous step. 

\paragraph{Projected Gradient Descent}\cite{PGD} then built on FGSM by introducing an iterative method of gradient descent, normalizing the outputs to limit the projection to a $L_p$ norm bound each step.

\paragraph{Carlini-Wagner}\cite{CW} later reformulated the task as a min-max optimization problem: how to maximize the change in the output of the system while minimizing the input variation to a $L_p$ norm.

\subsection{Adversarial attack detection and classification}
Adversarial attack detection is the task of binary classification between benign and adversarial attacks. This countermeasure can be thought of as a filter or a gatekeeper i.e. allowing only benign signals to pass. On the other hand, adversarial attack classification is a task of multiclass classification i.e. the problem of identifying the specific attack. While multiple works \cite{li2020investigating,AEDetection,peng2021pairing,chen2023detection} have been carried out for detection task, the classification task has been addressed by just a few papers\cite{villalba2020x,joshi2022advest}. Both papers used an x-vector network for representation learning to classify adversarial attacks. While \cite{villalba2020x} used the adversarial example directly, \cite{joshi2022advest} proposed using a denoiser to extract the adversarial perturbation and used it for classification. The denoiser used was a conditional GAN (CGAN)~\cite{mirza2014conditional,goodfellow2014generative} generative model with Multi-Resolution Short-time Fourier Transform (MRSTFT) loss~\cite{yamamoto2020parallel}.

The previous works in adversarial attack detector focus usually on thresholding techniques. The authors of \cite{AEDetection} extended the ParallelWaveGAN Vocoder used for preprocessing in ~\cite{joshi2021preprocessing} and used the difference between the resynthesized and actual scores to construct a detector for speaker verification systems. \cite{li2020investigating} proposed using a VGG-like network as a binary classification detector for speaker verification.
~\cite{peng2021pairing} proposed
a detector using 
two speaker verification models with a robustness gap (one model is a state-of-the-art model known to the adversary and not robust to adversarial attacks, while the other model is not known to the attacker and relatively robust against attack. This gap leads to score inconsistency that can be detected by a simple one-class classifier - minimum covariance determinant (MCD). ~\cite{chen2023detection} proposed  Minimum Energy in High FrEquencies for Short Time (MEH-FEST) detector that can detect FakeBob~\cite{chen2021real} attacks.

\label{subsec:previous_models}

\section{Dataset}
\label{sec:datasets}
In this paper, we use for all our experiments the \textit{Voxceleb1}~\cite{Nagrani17} and \textit{VoxCeleb2}~\cite{chung2018voxceleb2} datasets, which contain respectively the voices of a total of 1,251 and 5,996 celebrities, used generally for training and testing speaker verification systems.
The composition of each set is detailed in Table \ref{tab:voxceleb}
\begin{table}[ht]
    \centering
    \caption{VoxCeleb1 and 2 datasets distribution.}
    \begin{tabular}{l r r}
    \toprule
    \textbf{Set}             & \textbf{\# Speakers}    & \textbf{\# utterances} \\
    \midrule
    \textit{VoxCeleb1-dev}   & 1,211         & 148,642 \\
    \textit{VoxCeleb1-test}  & 40            & 4,874 \\
    \textit{VoxCeleb2-dev}   & 5,994         & 1,092,009 \\
    \bottomrule
    \end{tabular}
    
    \label{tab:voxceleb}
\end{table}
Those utterances are recorded from various multimedia sources, and collected on YouTube, with speakers mainly from the U.S, U.K, Germany, India, and France, speaking English for the most part.

We use \textit{VoxCeleb2-dev} to train all the victim models on a speaker identification task.
Those victim models are then evaluated for speaker verification, using three partitions of the \textit{VoxCeleb1-test} set: the \textit{Original} (O) partition, the \textit{Extended} (E) and the \textit{Hard} (H). Usually in speaker verification experiments, The \textit{VoxCeleb2-dev} set is used for training, and \textit{VoxCeleb1} is used for testing as there is no overlap of speakers. However, because we are evaluating for speaker recognition task, we would need overlapping speakers. Hence, we split the  \textit{VoxCeleb2-dev} into two parts \textit{VoxCeleb2-dev-train} and \textit{VoxCeleb2-dev-test}. We ensured that this split is 90\% train-10\% test per speaker. The \textit{VoxCeleb1-dev-train} set is used for training and validation of speaker recognition victim models and 
\textit{VoxCeleb2-dev-test} is used as the test set. 


\section{Threat Models}
\label{sec:threat_model}
This section presents the various threat models that are considered in this paper. We consider three untargeted white-box adversarial attacks (PGD, FGSM, and CW) using various $L_p$ norms ($p\in \{0,1,2,\infty\}$), against a set of speaker identification systems. Since the systems are under attack, they are called \textit{Victim models}.

\subsection{Victim Models}
\label{subsec:victim}
For the architecture of our victim models, we consider a set of four models:
\begin{itemize}
    \item ResNet34\cite{villalba2020state,he2016deep} (\textit{resnet34})
    \item LightResNet34\cite{zeinali2019but,villalba2020x} (\textit{lresnet34})
    \item Frequency-wise Squeeze Excitation ResNet34\cite{thienpondt2021integrating} (\textit{fwseresnet})
    \item ECAPA-TDNN\cite{desplanques2020ecapa} (\textit{ecapatdnn})
\end{itemize}
For reference, when the same models are used for speaker verification and trained on Voxceleb2-dev\cite{chung2018voxceleb2} and tested on all partitions of VoxCeleb1-test~\cite{Nagrani17}, the performances in Equal Error Rate (EER) of those models, evaluated on Voxceleb1 (Original, Extended, and Hard) are presented in Table \ref{tab:victim_models_perfs}

\begin{table}[ht]
    \centering
     \caption{Performances in EER of the victim models evaluated on Voxceleb1-O, Voxceleb1-H and Voxceleb1-E, using a Cosine distance.}
    \begin{tabular}{l c c c}
    \toprule
    \textbf{Model}       &    & $\downarrow$ \textbf{EER}   &   \\
                & \textit{Voxceleb1-O }& \textit{Voxceleb1-E}   & \textit{Voxceleb1-H}  \\
    \midrule
    resnet34    & 0.68\% & 0.86\% & 1.56\% \\
    lresnet34   & 1.59\% & 1.69\% & 2.84\% \\
    fwseresnet  & 0.73\% & 0.80\% & 1.58\% \\
    ecapatdnn   & 0.62\% & 0.83\% & 1.65\% \\
    \bottomrule
    \end{tabular}
   
    \label{tab:victim_models_perfs}
\end{table}

\subsection{Single Victim Model Dataset}
\begin{figure}[ht]
    \centering
    \includegraphics[width=0.5\textwidth]{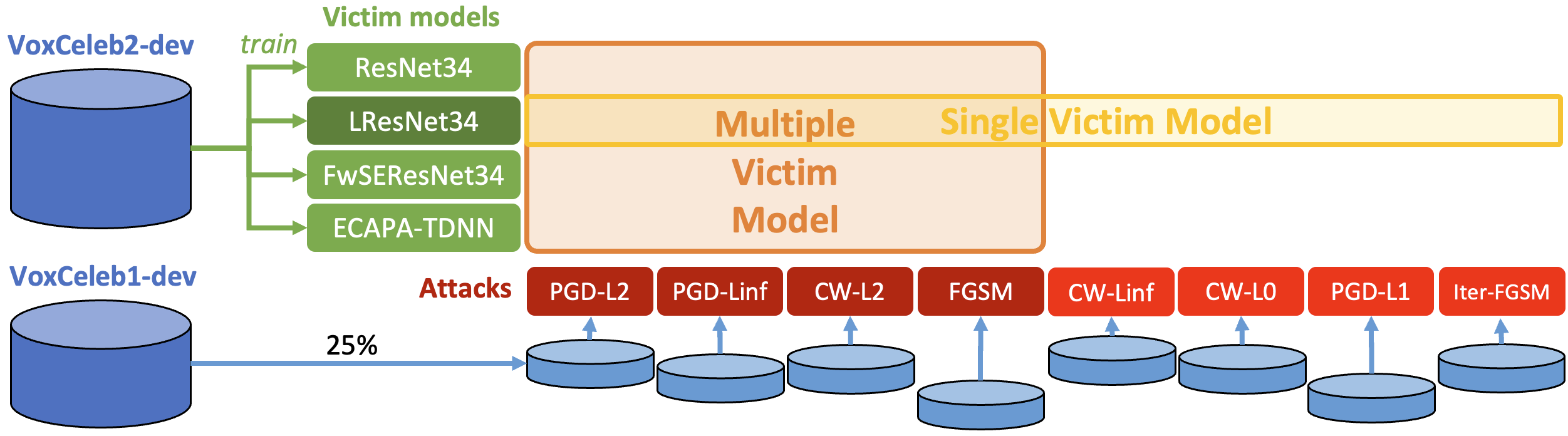}
    \caption{Schematic of the computation of the \textit{SingleVM} and the \textit{MultiVM} datasets.}
    \label{fig:threat_schematics}
\end{figure}
Our first dataset, the Single Victim Model (SingleVM) follows the same recipe as the one used in \cite{joshi2022advest}.
 The original VoxCeleb2 development set was split into two parts in the ratio 90\% to 10\% per speaker. We attack victim model \textit{lresnet34}, on a wide range of attacks (CW, FGSM, FGSM iterative, and PGD), with various norms ($L_0$, $L_1$, $L_2$, and $L_{\infty}$). For each of those attacks, we select a random 25\% split of the VoxCeleb1-dev dataset, compute the attack against the victim model, and keep the attacks that succeeded. The signature  victim models are trained using \textit{VoxCeleb2-dev-train}, and \textit{VoxCeleb2-dev-test} is used to test speaker classification tasks. We keep the associated benign utterances for each successful attack, which yields an additional 108,000 utterances. The test set is kept seperate. The distribution of utterances per attack is presented in Table \ref{tab:SingleVM_dataset}. 

\setlength{\tabcolsep}{1pt}
\begin{table}[ht]
    \centering
\captionsetup{justification=centering}
\caption{Distribution of the attacks generated for the Single Victim Model (SingleVM) dataset, train and test splits. The victim model used is always the Light ResNet34 here.}
    \resizebox{\columnwidth}{!}{
    \begin{tabular}{l c c c c}
    \toprule
    \textbf{Attack}  & \textbf{Norm}  & \textbf{Abbrev}. &  \textbf{SingleVM-train} & \textbf{SingleVM-test} \\
    \midrule
    CW      & $L_0$ & cw-l0 & 31,000    & 1679 \\
    CW      & $L_2$ & cw-l2 & 34,000    & 1774 \\
    CW      & $L_{\infty}$& cw-linf & 34,000 & 1774 \\
    FGSM    & $L_2$ & fgsm & 24,000    & 1429 \\
    Iterative FGSM & $L_2$ & iter-fgsm & 18,000 & 1034 \\
    PGD     & $L_1$ & pgd-l1 & 16,000    & 916 \\
    PGD     & $L_2$ & pgd-l2 & 16,000    & 895 \\
    PGD     & $L_{\infty}$ & pgd-linf & 17,000 & 966 \\
    \midrule
    All attacks & &   & 190,000   & 8,788 \\
    All benign  &  &  & 108,000   & 5,669 \\
    \midrule
    Total    &   &   & 298,000   & 14,457 \\
    \bottomrule
    \end{tabular}
    }
\label{tab:SingleVM_dataset}
\end{table}

\subsection{Multiple Victim Model Dataset}
As one victim model would not be enough to classify \textit{between} victim models, we generate a second dataset, the Multiple Victim Model dataset (\textit{MultiVM}), against the 4 victim models presented in Section \ref{subsec:victim}. 
We only compute the following 4 attacks: CW with $L_2$ norm, PGD with $L_2$ norm and $L_{\infty}$ norm, and FGSM to limit the size of the dataset and experimentation time, however, this dataset can be easily extended to include all the norms as in SingleVM.

This new dataset is computed the same way as the SingleVM but for each of the victim models. It is ensured that adversarial examples corresponding to benign utterances used for training and validation of the victim network are kept for training and validation of the attack classifier network. The test set is kept separate. The number of utterances for the generated dataset split into the train, validation, and test sets is shown in Table \ref{tab:MultiVM_dataset}.
Figure \ref{fig:threat_schematics} illustrates the making of both threat model datasets from  Voxceleb2-dev set.



\begin{figure}[ht]
    \centering
\captionsetup{justification=centering}
    \includegraphics[width=0.5\textwidth]{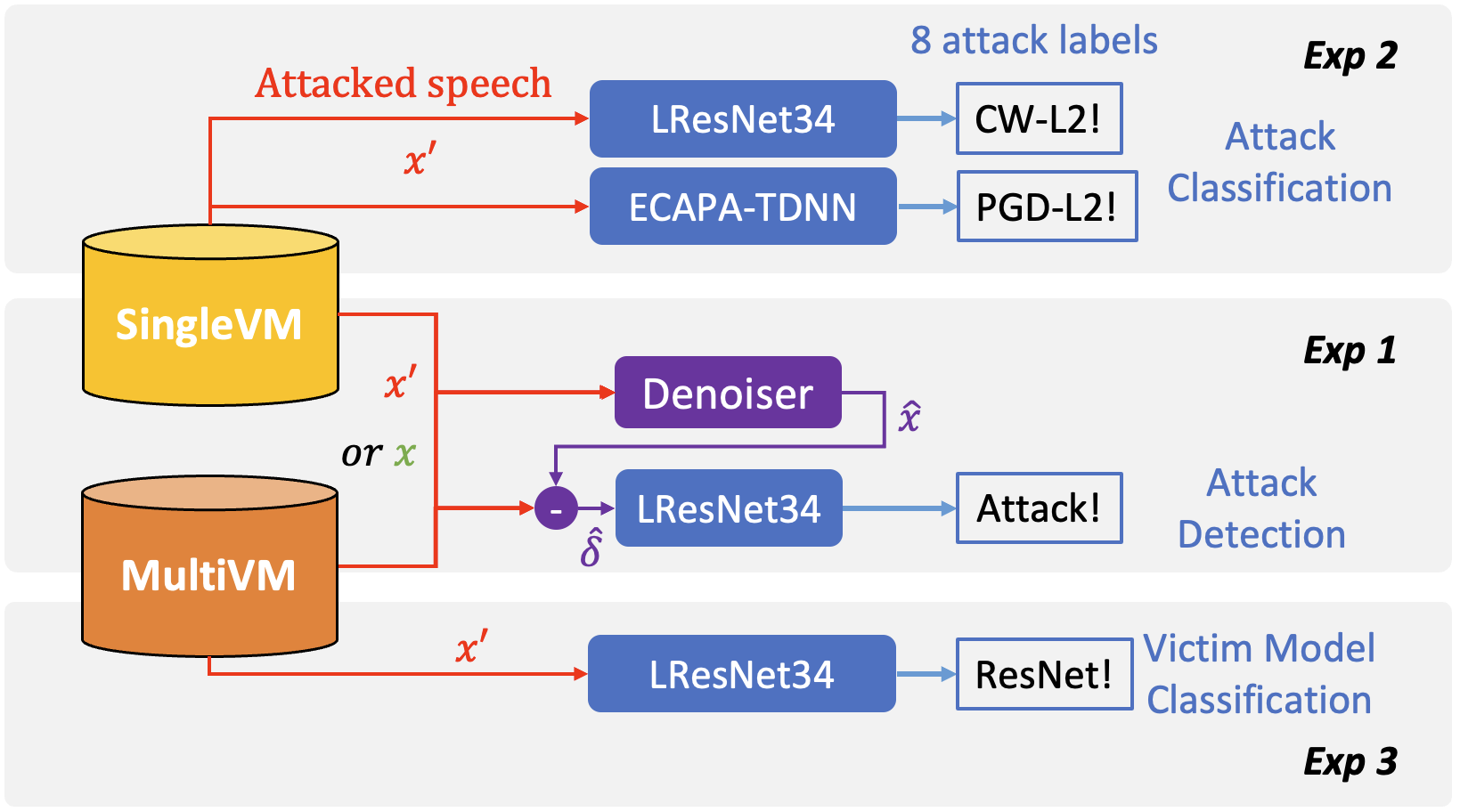}
    \caption{Pipeline of inference to predict the label linked to a given utterance $x'$ of the \textit{SingleVM} or the \textit{MultiVM} dataset, for the three experiments proposed.}
    \label{fig:pipeline}
\end{figure}
\begin{figure}
    \centering
\captionsetup{justification=centering}
\includegraphics[width=0.5\textwidth]{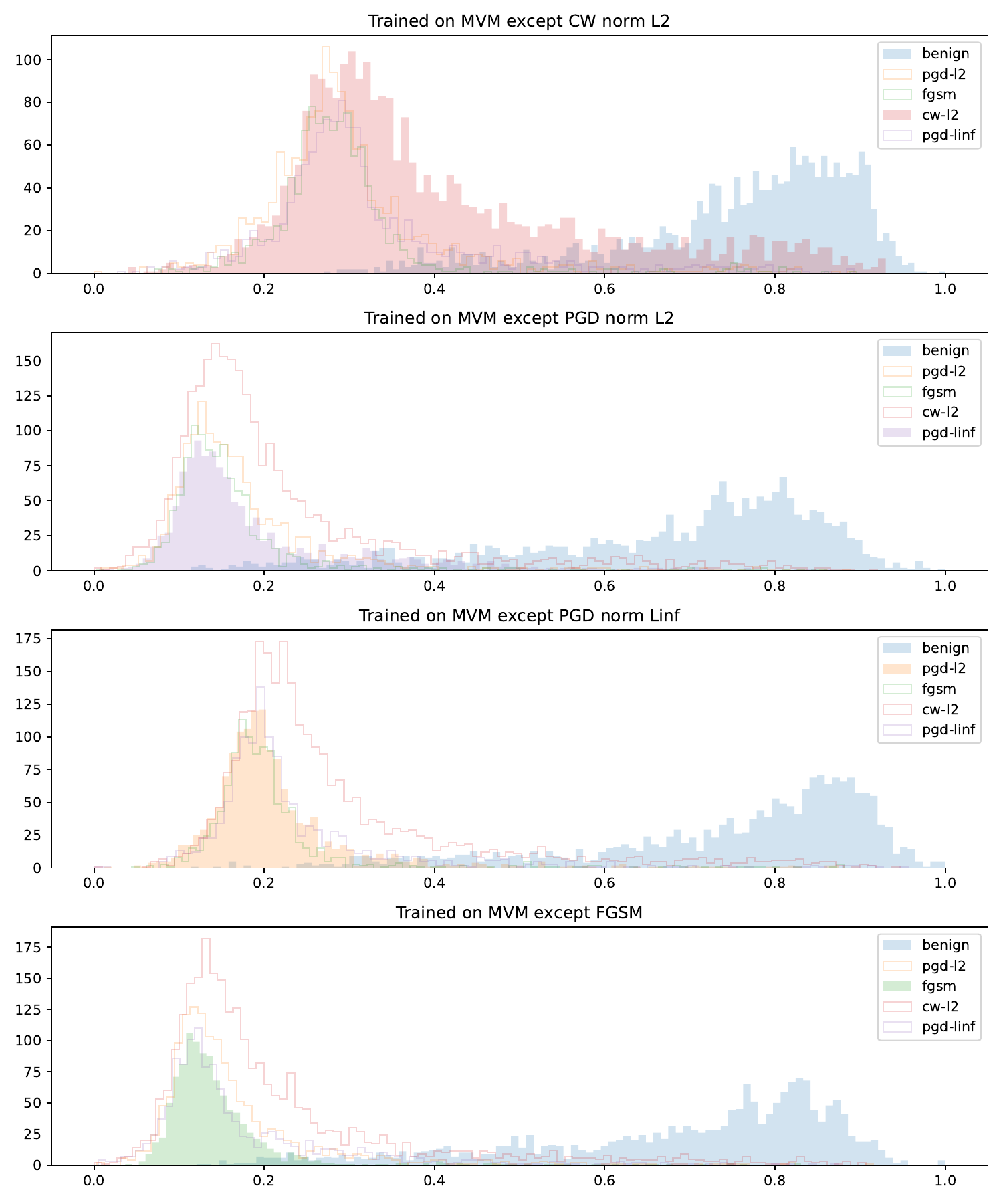}
    \caption{Score distribution of the \textit{MultiVM-test} data, for each system trained on all but one attack. The benign utterances are always in filled light blue, the attack that was removed from the train set is filled while the others are emptied.}
    \label{fig:binary_ablation_scores}
\end{figure}
\begin{table}[ht]
\centering
\captionsetup{justification=centering}
        \caption{Data distribution for the Multiple Victim Model (MultiVM) dataset. Numbers indicate the number of utterances. The total dataset has 127,261 utterances across the four victim models.}
    \resizebox{0.7\columnwidth}{!}{
    \begin{tabular}{@{}c|cccccccc@{}}
    \toprule
     & \textbf{Train Set} & \textbf{Validation Set}	& \textbf{Test Set} \\
     \midrule
    ecapatdnn	& 27,149	& 3,232	 & 1,582  \\
    fwseresnet	& 25,514	& 3,026	& 1,521  \\
    resnet	& 25,715	& 3,064	& 1,522  \\
    lresnet	& 29,719	& 3,514	& 1,703  \\
    \midrule
    \textit{Total}	& \textit{108,097} & 	\textit{12,836}	& \textit{6,328}  \\
    \bottomrule
    \end{tabular}
    \vspace{-1cm}
    }
\label{tab:MultiVM_dataset}
\end{table}


\begin{table*}[t]
    \centering
\captionsetup{justification=centering}
 \caption{AUC, Accuracy and EER of our various attack detection experiments, for the Single Victim Model dataset (\textit{SingleVM}), for the Multiple Victim Model Dataset (\textit{MultiVM}), and our ablation studies on the Multiple Victim Model Dataset, using all but one attack, evaluated on the whole dataset, on all but the attack, and only on the benign and the attack.}
    \begin{tabular}{l l c c c}
    \toprule
    \textbf{Trained on} & \textbf{Tested on}  &   $\uparrow$ \textbf{AUC} &  $\uparrow$ \textbf{Accuracy}    & $\downarrow$ \textbf{EER} \\
    \midrule
    \textit{SingleVM-train}       &   \textit{SingleVM-test}                   & 0.982 & 93.7\% & 6.34\% \\
    \midrule
    \midrule
    \textit{MultiVM-train}      &   \textit{MultiVM-test}           & 0.961 & 91.1\% & 8.89\% \\
    \midrule
                &   \textit{MultiVM-test}                      & 0.953 & 89.3\% & 10.67\% \\
    \textit{MultiVM-train} except cw-l2
                &  \textit{MultiVM-test}  except cw-l2      & 0.979 & 92.9\% & 7.13\% \\
                &  only benign and cw-l2   & 0.916 & 85.3\% & 14.68\% \\
    \midrule
                &   \textit{MultiVM-test}                        & 0.968 & 91.9\% & 8.06\% \\
    \textit{MultiVM-train} except fgsm  
                &   \textit{MultiVM-test}  except fg    sm     & 0.965 & 91.4\% & 8.63\% \\
                &   only benign and fgsm      & 0.983 & 95.5\% & 4.46\% \\
    \midrule
                &   \textit{MultiVM-test}                       & 0.955 & 91.1\% & 8.94\% \\
    \textit{MultiVM-train} except pgd-l2
                & \textit{MultiVM-test}  except pgd-l2      & 0.954 & 91.3\% & 8.67\% \\
                & only benign and pgd-l2   & 0.956 & 90.5\% & 9.50\% \\
    \midrule
                &   \textit{MultiVM-test}                     & 0.967 & 92.1\% & 7.92\% \\
    \textit{MultiVM-train} except pgd-linf 
                & \textit{MultiVM-test}  except pgd-linf & 0.964 & 91.5\% & 8.51\% \\
                & only benign and pgd-linf & 0.980 & 94.9\% & 5.12\% \\
    \bottomrule
    \end{tabular}
    \label{tab:results_binary}
\end{table*}

\subsection{Attack classification models}
The attack classification networks were built with the same architectures, but the embedding dimension was empirically set to 10.  We explore two architectures:
\begin{itemize}
    \item \textbf{LightResNet34}, the same architecture as in \cite{joshi2022advest} and also one of the victim model used here.
    \item \textbf{ECAPA-TDNN}~\cite{desplanques2020ecapa}, a popular architecture using attention, to improve the performances of our model.
\end{itemize}

\section{Experiments}
\label{sec:experiments}

Our experiments follow a similar structure.
For a given sample $x'$ (that could be benign or adversarial), we can first use a pre-trained denoiser $D$ (same a \cite{joshi2022advest} and as briefly described in Section \ref{subsec:previous_models}) to predict an estimation of the clean associated sample $\hat{x}$, which is then used to estimate the adversarial noise $\hat{\delta}$.
We then train a classifier $C$ to predict a given class either from the estimated adversarial noise $\hat{\delta}$ or directly from the attacked signal $x'$. Our experiments  explore three directions:
\begin{enumerate}
  \item What are the performances of our pipeline when used only for attack detection (as a binary classifier)?
    \item Does the architecture of the classifier play a role in the attack classification?
    \item Can we classify the victim model on which the attacks were computed?
\end{enumerate}
The following subsections delve into these directions. All experiments are performed using the Hyperion toolkit\footnote{https://github.com/hyperion-ml/hyperion}.

\subsection{Attack Detection}
This experiment explores the potential of the best of our classifiers to be retrained for a binary task: attack vs benign classification.
We train a LightResNet34 on our Single Victim Model dataset, using a pre-trained denoiser to estimate the adversarial noise $\hat{\delta}$, and find out the best performances we could get on 8 different attacks.
Then, we train the same architecture on our Multiple Victim Model dataset to compare the results on a comparatively limited set of attacks. We chose these attacks because they represent the family well and we wanted to avoid confusion between the norms to see whether the model learns the underlying victim models better.

We also study the impact of having one unknown attack, by training 4 classifiers, each ignoring one of the four attacks from our Multiple Victim Model dataset, and look at the performance classifications on the unknown attack.
The performances are evaluated in terms of Area Under the Curve (AUC), Accuracy, and Equal Error Rate (EER) and presented in Section \ref{subsec:results_binary}

\subsection{Different architectures for classifier}
Our first experiment explores changes in architectures for the classifier, facing an attack that was always performed against the same victim model, a LightResNet34.
Because we want to compare the impact of a different classifier, we choose to not use the denoiser to compare performances here, as we don't want any interference in the process.

We compare two architectures with our Single Victim Model dataset:
\begin{itemize}
    \item \textbf{LightResNet34}, the same architecture as in \cite{joshi2022advest} and also the one of the victim model used here.
    \item \textbf{ECAPA-TDNN}~\cite{desplanques2020ecapa}, a popular architecture using attention, to improve the performances of our model.
\end{itemize}
For both experiments, we train our classifier using the train split of our Single Victim Model dataset, to predict the attack of each utterance.The accuracy of attack classification is detailed and discussed in Section \ref{subsec:results_archis}.

\subsection{Victim Model Classification}
Our third experiment targets the classification of the victim model on which the adversarial attack was computed.
We train a LightResNet34 classifier on the adversarial utterances of our Multiple Victim Model dataset to classify between the 4 victim models.  The accuracy of victim model classification is detailed and discussed in Section \ref{subsec:results_model_classif}.

\section{Results}
\label{sec:results}
This section presents the results of our three experiments along with their associated conclusions.

\subsection{Attack Detection}
\label{subsec:results_binary}
Table \ref{tab:results_binary} presents the results in AUC, Accuracy and EER of our various attack detection experiments.

Table \ref{tab:results_binary} shows good classification results on the \textit{SingleVM} dataset, but a significant degradation once trained the \textit{MultiVM}, which is understandable as the latest only contains a subset of attacks and has comparatively less data.
We can also see that removing one of the attacks present only marginally impacts the binary classification results.

To further understand how the ablation impacts the distribution of the scores, we plot the distribution of the scores for the \textit{MultiVM} dataset in each of the cases in Figure \ref{fig:binary_ablation_scores}.

We can see in Figure \ref{fig:binary_ablation_scores} score distribution of the \textit{MultiVM-test} data, for each system trained on all but one attack. It can be seen that removing one attack has a low effect on the general distribution of the scores.

\begin{figure}[t]
    \centering
\captionsetup{justification=centering}
\includegraphics[width=0.55\textwidth]{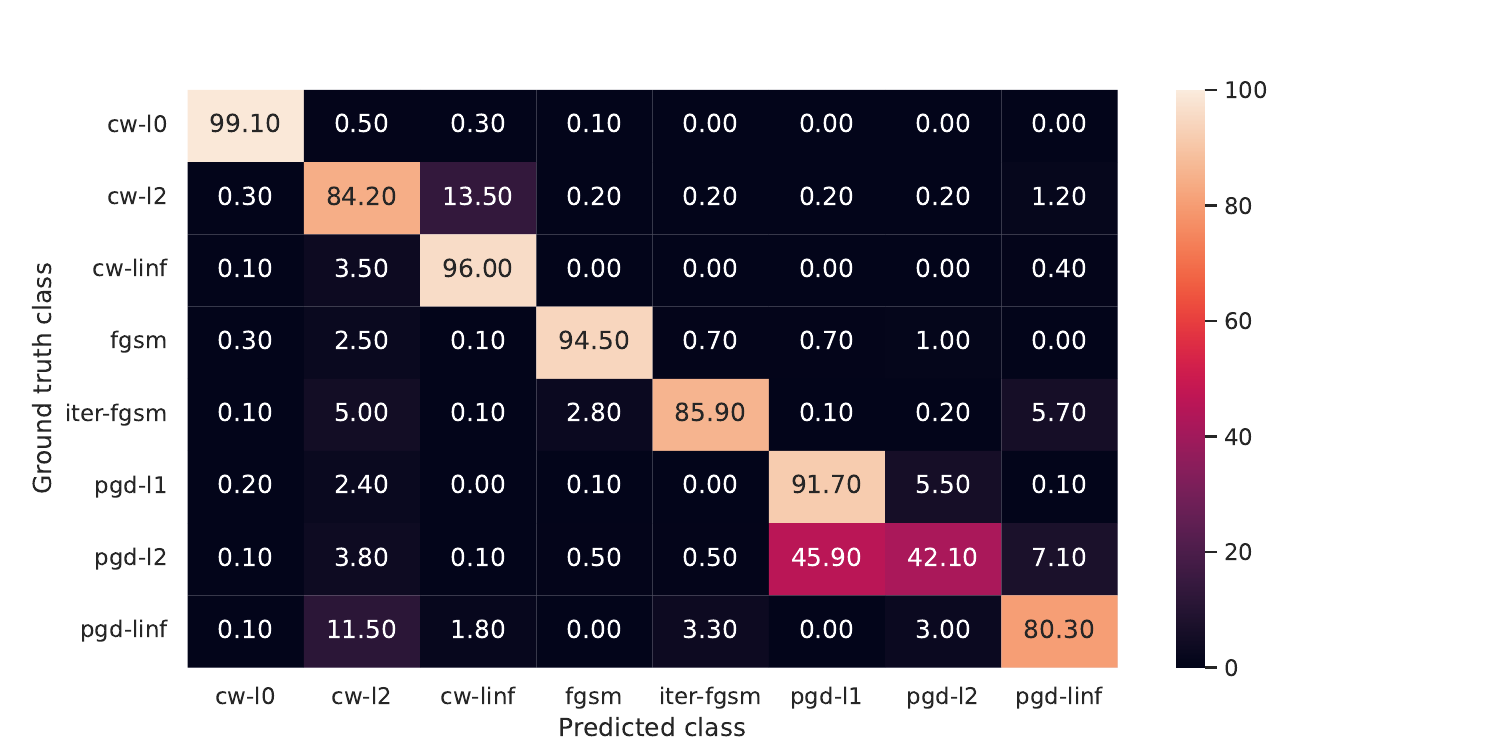}
    \caption{Normalized confusion matrix (\%) using LResNet34 as classifier for SingleVM dataset}
    \label{fig:cm_lresnet}
\end{figure}

\begin{figure}[ht]
    \centering
\captionsetup{justification=centering}
\includegraphics[width=0.55\textwidth]{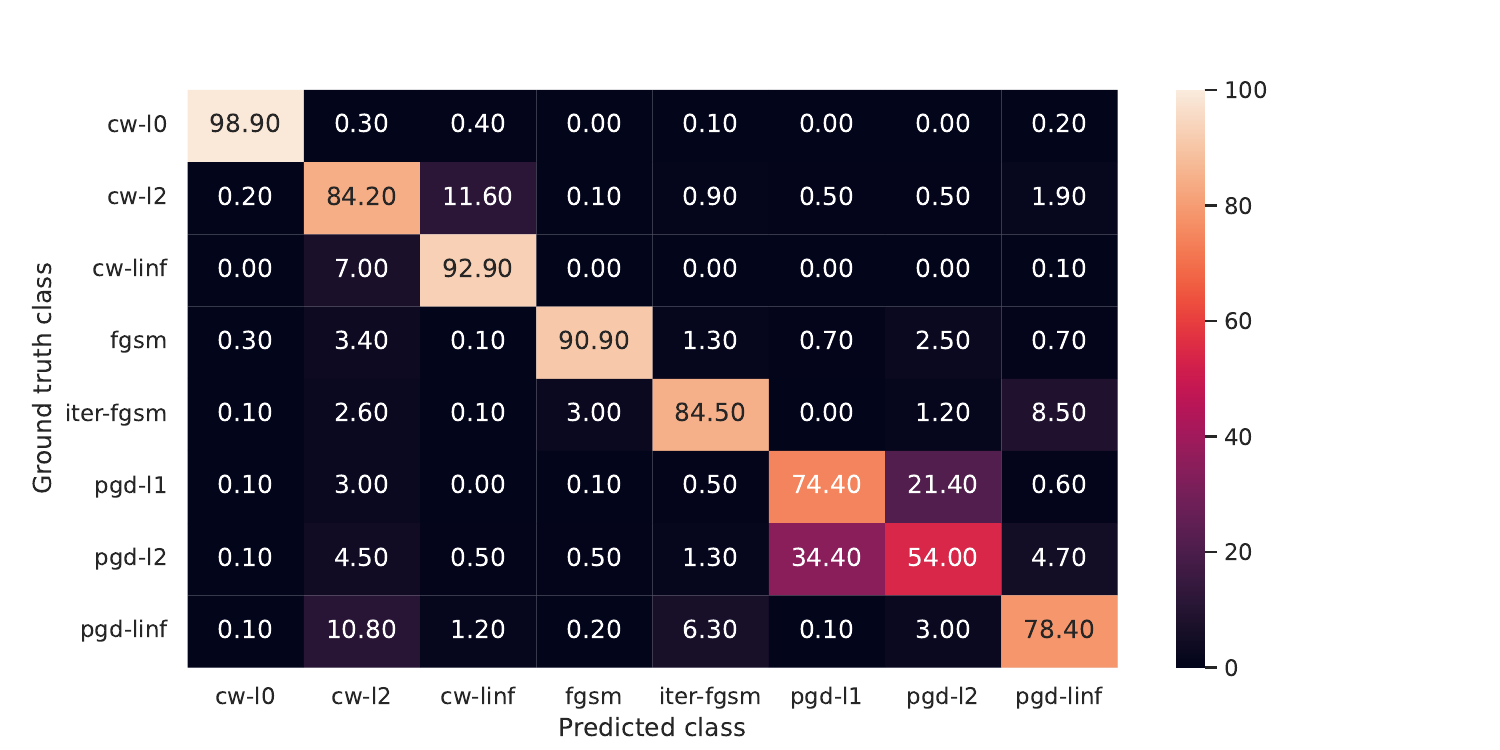}
    \caption{Normalized confusion matrix (\%) using ECAPA-TDNN as classifier for SingleVM dataset}
    \label{fig:cm_ecapatdnn}
\end{figure}
\subsection{Different architectures for classifier}
\label{subsec:results_archis}
\begin{figure}[ht]
    \centering
\captionsetup{justification=centering}
    \includegraphics[width=0.5\textwidth]{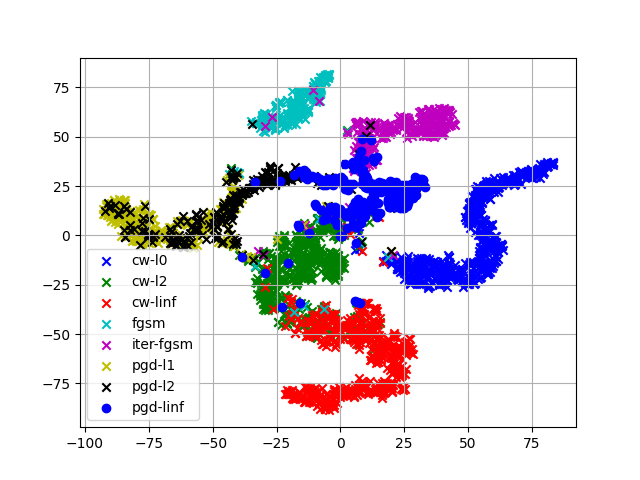}
    \vspace{-1cm}
    \caption{TSNE for LResNet34 architecture as attack classifier on \textit{SingleVM} dataset}
    \label{fig:tsne_lresnet}
\end{figure}

\begin{figure}[ht]
\captionsetup{justification=centering}
    \vspace{-0.5cm}
    \centering
    \includegraphics[width=0.5\textwidth]{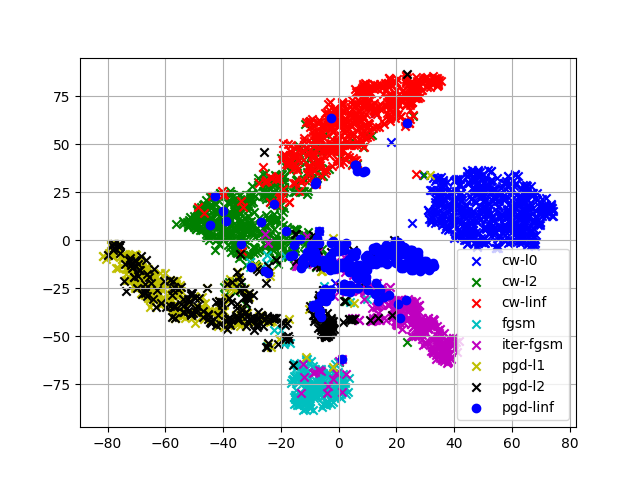}
    \vspace{-1cm}
    \caption{TSNE for ECAPA-TDNN architecture as attack classifier on \textit{SingleVM} dataset}
    \label{fig:tsne_ecapa}
\end{figure}

We observe significant differences in classification performance when different architectures are used for attack classification for the same victim model. We use \textit{SingleVM} dataset and vary the architecture for the classifier network  (LResNet, ECAPA-TDNN) as described in the following sections. The overall classification accuracy for LResNet is 86.48 \% and the confusion matrix is shown in Figure \ref{fig:cm_lresnet}. The overall classification accuracy for ECAPA-TDNN is 84.79\% and the confusion matrix is shown in Figure \ref{fig:cm_ecapatdnn}
The overall accuracy of the two classifiers does not differ by a very significant margin (absolute difference of 1.69\%). We compared the correctly classified accuracy (diagonals of the confusion matrices) of the two architectures and plotted them in Table \ref{tab:comparison}. The accuracies are quite close to each other and ECAPA-TDNN is only better for PGD-L2 attack which can also be seen from comparing the TSNE figures Figure \ref{fig:tsne_lresnet} for LResNet and Figure \ref{fig:tsne_ecapa}. We can see in both matrices that the main confusion is always between PGD $L_1$ and PGD $L_2$ attacks.
\begin{table}[ht]
\captionsetup{justification=centering}
\caption{Comparison between the two classifiers for victim model classification in speaker recognition.}
    \centering
    \label{tab:comparison} 
    \resizebox{0.5\columnwidth}{!}{
    \begin{tabular}{@{}c cc@{}}
    \toprule
    	& \textbf{lresnet}	& \textbf{ecapatdnn} \\
     \midrule
    \textbf{cw-l0}		& 99.1	& 	98.9\\
   \textbf{ cw-l2}		& 84.2		& 84.2\\
    \textbf{cw-linf	}	& 96		& 92.9\\
    \textbf{fgsm}		& 94.5		& 90.9\\
    \textbf{iter-fgsm}		& 85.9		& 84.5\\
    \textbf{pgd-l1}		& 91.7		& 74.4\\
    \textbf{pgd-l2}		& 42.1		& \textbf{54.0}\\
    \textbf{pgd-linf}		& 80.3	& 	78.4\\
    \bottomrule
    \end{tabular}
    }
\end{table}

\subsection{Victim Model Classification}
\label{subsec:results_model_classif}

\begin{figure}
\captionsetup{justification=centering}
\includegraphics[width=0.5\textwidth]{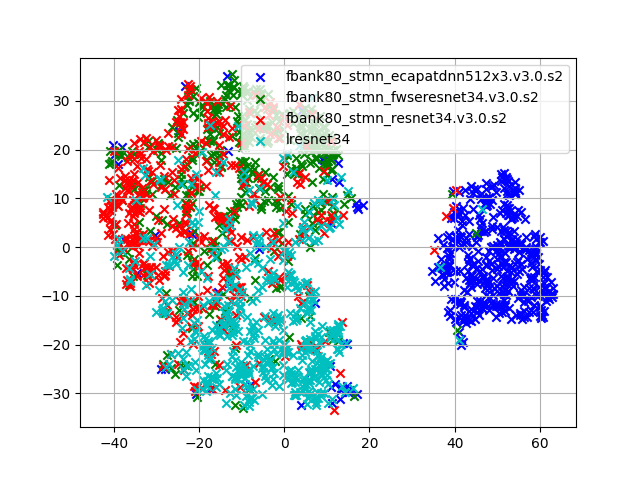}
    \vspace{-1cm}
  \caption{TSNE for victim model classification on \textit{MultiVM} dataset}
  \label{fig:tsne_victim_model}
\end{figure}

For the victim model classification, we obtain a classification accuracy of 72.28\% for the 4 models of \textit{MultiVM} test. 
The confusion matrix is shown in Table \ref{tab:class_victim_model} and the corresponding TSNE plot is in Figure \ref{fig:tsne_victim_model}. 
It can be clearly seen that the adversarial attacks on \textit{ecapatdnn} can be detected with the highest accuracy. 
The corresponding TSNE also shows the clear separation of ECAPA-TDNN embeddings from the rest of the embeddings. We suspect this is because the architecture of \textit{ecapatdnn} is very different from the other three (\textit{fwseresnet, resnet34, lresnet34}) which fall under the same broad category of the ResNet34.

\setlength{\tabcolsep}{2pt}
\begin{table}[ht]
\captionsetup{justification=centering}
\caption{Normalized confusion matrix (\%) for victim model classification in speaker recognition.}
    \label{tab:class_victim_model} 
    \resizebox{\columnwidth}{!}{
    \begin{tabular}{@{}c ccccccccc@{}}
    \toprule
     & \textbf{ecapatdnn} & \textbf{fwseresnet}	& \textbf{resnet34}	& \textbf{lresnet34} \\
     \midrule
    \textbf{ecapatdnn}	 & \bf 92.0     &  	1.9   & 	1.4   &   	4.7 \\
    \textbf{fwseresnet}  & 	2.9 & 	\bf 59.5 & 	23.9	 & 13.7 \\
    \textbf{resnet34}	 & 1.5  & 	19.3  & 	\bf 54.7  & 	24.5 \\
    \textbf{lresnet34}	 &  1.5 & 	5.3  & 	12.2	 & \bf 81.1 \\
    \bottomrule
    \end{tabular}
    }
\end{table}

\section{Conclusion}
\label{sec:conclusion}



In this study, we have undertaken a thorough examination of adversarial attacks targeting speaker verification systems, aiming to fortify the resilience of such systems against malicious intrusions. 
Our investigation builds upon existing defense and classification methodologies while innovatively extending them to encompass both attack detection and classification of adversarial attacks and victim models.

Our findings demonstrate the robustness and effectiveness of our framework, showcasing its capability to accurately detect the presence of adversarial attacks with an impressive AUC of up to 0.982. 
Furthermore, our classification accuracy of 86.48\% over eight classes for the LResNet34 classifier underscores the efficacy of our approach in discerning the type of attack employed. Notably, the introduction of our novel handcrafted dataset, comprising various attacks inflicted upon different victim models, has significantly enhanced our ability to detect and classify a broader spectrum of attacks, achieving a commendable classification accuracy of 72.28\% over four classes.

Looking forward, our research opens up promising avenues for further exploration and advancement. 
Refinements and optimizations of our defense framework hold the potential to bolster its effectiveness in real-world scenarios, particularly by broadening the scope to encompass a wider array of victim models and diverse attack parameters. 
Moreover, we aspire to expand our repertoire to include black-box attacks and adaptive attacks, thereby enriching the diversity of threat models considered and fortifying our defenses against an even broader spectrum of adversarial intrusions. 
By continuously evolving and refining our methodologies, we strive to propel the field of adversarial defense forward, ensuring the robustness and reliability of speaker identification systems in the face of evolving security threats.

\section{Acknowledgments}
This research has been supported by DARPA RED \url{https://www.darpa.mil/program/reverse-engineering-of-deceptions} under contract HR00112090132 

\bibliographystyle{IEEEbib}
\bibliography{bibliography}
\end{document}